\documentclass[aps,preprint,showpacs]{revtex4-1}
\usepackage{amsmath}
\usepackage{amsfonts}
\usepackage{amssymb}
\usepackage{graphicx}
\usepackage{bm}
\begin{document}
\title{Quantum beats of the rigid rotor}
\author{K. Kowalski and J. Rembieli\'nski}
\affiliation{Department of Theoretical Physics, University
of \L\'od\'z, ul.\ Pomorska 149/153, 90-236 \L\'od\'z,
Poland}
\begin{abstract}
The dynamics is investigated of a free particle on a sphere (rigid rotor or
rotator) that is initially in a coherent state.  The instability of coherent
states with respect to the free evolution leads to nontrivial time-development 
of averages of observables representing the position of a particle on a sphere
that can be interpreted as quantum beats.
\end{abstract}
\maketitle
\section{Introduction}
The rigid rotor model is a very important concept in quantum mechanics that is 
discussed in most textbooks.  We only recall that it is basic for understanding
rotational spectroscopy and collisions of molecules.  As is well known a 
mathematical model for rigid rotor is a free particle constrained to a surface
of a sphere \cite{1}.  In this work we study the quantum dynamics of a 
free particle on a sphere in the case when the initial condition is a coherent
state.  The coherent states for the quantum mechanics on a sphere are unstable
with respect to the free evolution, nevertheless, in opposition to the motion
in a plane, the corresponding wave packets do not spread.  Consequently, the
dynamics is nontrivial.  In particular, the evolution of the wave packets leads
to beats in the expectation values of position observables for a particle on a 
sphere. 
\section{Quantum mechanics on a sphere}
We begin with a brief account of the quantum mechanics on a sphere.  As shown in
\cite{2} the most natural algebra for the study of the motion on a sphere $S^2$
is the $e(3)$ algebra of the form
\begin{equation}
[J_i,J_j]={\rm i}\varepsilon_{ijk}J_k,\quad [J_i,X_j]={\rm i}
\varepsilon_{ijk}X_k,\quad [X_i,X_j]=0,
\end{equation}
where ${\bm J}$ is the angular momentum operator and ${\bm X}$ is the position 
operator for a particle on a sphere.  In fact, the algebra (1) has two Casimir 
operators given in a unitary irreducible representation by
\begin{equation}
{\bm X}^2=r^2,\qquad {\bm J}{\bm\cdot}{\bm X}=\lambda.
\end{equation}
In the following we restrict to the irreducible representations of (1) with $r=1$
i.e.\ the unit sphere and the ``most classical" case of $\lambda=0$.  Then the angular
momentum basis spanned by the common eigenvectors of operators ${\bm J}^2$, $J_3$, 
${\bm X}^2$ and ${\bm J}{\bm\cdot}{\bm X}$ is defined by
\begin{subequations}
\begin{eqnarray}
&&{\bm J}^2 |j,m\rangle = j(j+1) |j,m\rangle,\quad J_3
|j,m\rangle=m|j,m\rangle,\\
&&{\bm X}^2 |j,m\rangle=|j,m\rangle,\qquad
({\bm J}{\bm\cdot}{\bm X}) |j,m\rangle=0,
\end{eqnarray}
\end{subequations}
where $j$ is a nonnegative integer and $-j\le m\le j$.  Besides of the standard action
of the angular momentum operators $J_\pm=J_1\pm{\rm i}J_2$ on the basis vectors 
$|j,m\rangle$ such that
\begin{equation}
J_\pm |j,m\rangle=\sqrt{(j\mp m)(j\pm m+1)}\,|j,m\pm 1\rangle
\end{equation} 
we have the following relations satisfied by the position operators \cite{2}
\begin{subequations}
\begin{eqnarray}
X_\pm |j,m\rangle
&=&\mp\frac{\sqrt{(j\pm m+1)(j\pm m+2)}}
{\sqrt{(2j+1)(2j+3)}}|j+1,m\pm1\rangle\nonumber\\
&&{}\pm\frac{\sqrt{(j\mp m-1)(j\mp m)}}{j\sqrt{(2j-1)(2j+1)}}
|j-1,m\pm1\rangle,\\
X_3 |j,m\rangle
&=&\frac{\sqrt{(j-m+1)(j+m+1)}}
{\sqrt{(2j+1)(2j+3)}}|j+1,m\rangle\nonumber\\
&&{}+\frac{\sqrt{(j-m)(j+m)}}{\sqrt{(2j-1)(2j+1)}}|j-1,m\rangle,
\end{eqnarray}
\end{subequations}
where $X_\pm=X_1\pm {\rm i}X_2$.  We finally write down the orthogonality 
relations satisfied by the vectors $|j,m\rangle$ such that
\begin{equation}
\langle j,m|j',m'\rangle=\delta_{jj'}\delta_{mm'},
\end{equation}
and the completeness condition which in the case of $\lambda=0$ takes the form
\begin{equation}
\sum_{j=0}^{\infty}\sum_{m=-j}^{j}
|j,m\rangle\langle j,m|=I.
\end{equation}
\section{Coherent states for the sphere}
We now summarize the basic facts about the coherent states for a particle on a 
sphere.  The coherent states for the quantum mechanics on a sphere $S^2$ were 
introduced by us very recently \cite{2}.  These states were generalized 
by Hall and Mitchell \cite{3} to the case involving $n$-dimensional sphere.  For 
$n=1$ they reduce to the coherent states for a particle on a circle $S^1$
introduced by us in the paper \cite{4} (see also \cite{5}) and utilized for the 
construction of coherent states on a torus \cite{6}.  An alternative construction 
of coherent states for the sphere $S^n$, $n\ge2$, was described by D\'{\i}az-Ortiz 
and Villegas-Blas in a very recent work \cite{7}.  The coherent states can be defined
as the solution of the eigenvalue equation \cite{2}
\begin{equation}
{\bm Z} |{\bm z}\rangle = {\bm z} |{\bm z}\rangle
\end{equation}
where ${\bm Z}$ is given by
\begin{eqnarray}
{\bm Z} &=&\left(\frac{e^{1/2}}{\sqrt{1+4{\bm J}^2}}{\rm
sinh}\hbox{$\scriptstyle 1\over2 $}\sqrt{1+4{\bm
J}^2}+e^{1/2}{\rm cosh}\hbox{$\scriptstyle 1\over2 $}
\sqrt{1+4{\bm J}^2}\right){\bm X}\nonumber\\
&&{}+{\rm i}\left(\frac{2e^{1/2}}{\sqrt{1+4{\bm J}^2}}{\rm sinh}
\hbox{$\scriptstyle 1\over2 $}\sqrt{1+4{\bm J}^2}\right){\bm
J}\times{\bm X},
\end{eqnarray}
where the cross designates the vector product.  The operator ${\bm Z}$ and
${\bm z}\in{{\mathbb C}^3}$ satisfy
\begin{equation}
{\bm Z}^2=1,\qquad {\bm z}^2=1.
\end{equation} 
The first equation of (10) is evident in view of the relation \cite{8}
\begin{equation}
{\bm Z} = e^{-{\bm J}^2/2}{\bm X}e^{{\bm J}^2/2}.
\end{equation}
The coherent states $|{\bm z}\rangle$ can be generated from the coherent
state $|{\bm e}_3\rangle$ (fiducial vector), where ${\bm e}_3=(0,0,1)$, 
via
\begin{equation}
|{\bm z}\rangle = \exp\left[\frac{{\rm arccosh}z_3}{\sqrt{1-z_3^2}}
({\bm z}\times{\bm e}_3){\bm\cdot}{\bm J}\right]
|{\bm e}_3\rangle,
\end{equation}
where
\begin{equation}
|{\bm e}_3\rangle=\sum_{j=0}^{\infty}e^{-\frac{1}{2}j(j+1)}\sqrt{2j+1}|j,0\rangle.
\end{equation}
The projection of the coherent state on the basis $|j,m\rangle$ is given by
\begin{widetext}
\begin{equation}
\langle j,m|{\bm z}\rangle = e^{-\frac{1}{2}j(j+1)}\sqrt{2j+1}\,
\frac{(2|m|)!}{|m|!}\sqrt{\frac{(j-|m|)!}{(j+|m|)!}}\left(
\frac{-\varepsilon(m)z_1+{\rm i}z_2}{2}\right)^{|m|} C_{j-|m|}^{|m|+\frac{1}{2}}
(z_3),
\end{equation}
\end{widetext}
where $C^\alpha_n(x)$ are the Gegenbauer polynomials and $\varepsilon(m)$ is the 
sign of $m$.  The coherent states are not orthogonal.  We have \cite{8}
\begin{equation}
\langle {\bm z}|{\bm w} \rangle =
\sum_{j=0}^{\infty}e^{-j(j+1)}(2j+1)P_j({\bm z}^*{\bm\cdot}{\bm w}),
\end{equation}
where $P_j(x)$ are the Legendre polynomials.

The natural parametrization of ${\bm z}$ by points of the classical phase space
is given by
\begin{equation}
{\bm z}=\cosh|{\bm l}|\,{\bm x}+{\rm i}(\sinh|{\bm l}|/
|{\bm l}|){\bm l}\times {\bm x},
\end{equation}
where the vectors ${\bm l},\,{\bm x}\in{\mathbb R}^3$, fulfil
${\bm x}^2=1$ and ${\bm l}{\bm\cdot}{\bm x}=0$, that is we assume that ${\bm l}$ 
is the classical angular momentum, $|{\bm l}|$ is the norm of the vector ${\bm l}$, 
and ${\bm x}$ is the radius vector of a particle on a unit sphere.  Introducing the 
spherical coordinates ${\bm x}=(\sin\theta\cos\varphi,\sin\theta\sin\varphi, 
\cos\theta)$, and parametrizing the tangent vector ${\bm l}$ by its norm $|{\bm l}|$ 
and the angle $\alpha$ between ${\bm l}$ and the meridian passing through the point 
with the radius vector ${\bm x}$, we get from (16) the following natural 
parametrization of the phase space compatible with the constraints
(see \cite{8}):
\begin{equation}
{\bm z} = \cosh|{\bm l}|\,{\bm x} + {\rm i}\sinh|{\bm l}|(\sin\alpha{\bm n}
+\cos\alpha{\bm n}_0),
\end{equation}
where ${\bm n}=(\cos\varphi\cos\theta,\sin\varphi\cos\theta,-\sin\theta)$ and 
${\bm n}_0=(-\sin\varphi,\cos\varphi,0)$ are the unit mutually orthogonal vectors.
The correctness of the introduced coherent states for a sphere is confirmed by the
good behavior of quantum averages.  Namely, we have 
\begin{equation}
\frac{\langle {\bm x},{\bm l}|{\bm J}|{\bm x},{\bm l}\rangle}{\langle 
{\bm x},{\bm l}|{\bm x},{\bm l}\rangle}\approx{\bm l} 
\end{equation}
where $|{\bm x},{\bm l}\rangle\equiv|{\bm z}\rangle$, with ${\bm z}$ given by (16),
and the approximation is very good.  From computer calculations it follows that  
for $|{\bm l}|\ge10$, the relative error is of order 1\%.  Furthermore, the computer
simulations indicate that
\begin{equation}
\frac{\langle {\bm x},{\bm l}|{\bm X}|{\bm x},{\bm l}\rangle}{\langle 
{\bm x},{\bm l}|{\bm x},{\bm l}\rangle}\approx e^{-1/4}{\bm x},
\end{equation}
where the approximation is as good as in (18).  We point out that the factor $e^{-1/4}$
in (19) is related to the fact that ${\bm X}$ is not diagonal in the coherent state
basis.  Proceeding analogously as in the case of the coherent states for a circle 
\cite{4} one can introduce relative expectation value $\langle\!\langle {\bm X}\rangle
\!\rangle$ of ${\bm X}$ with respect to averages in the coherent states labelled
by unit vectors ${\bm x}={\bm e}_i$, so that $\langle\!\langle {\bm X}\rangle
\!\rangle\approx{\bm X}$.  Yet another evidence of correctness of the described 
coherent states is behavior of the corresponding wave functions such that \cite{8}
\begin{equation}
f_{\bm z}({\bm x}) = \frac{1}{\sqrt{4\pi}}
\sum_{j=0}^{\infty}e^{-(1/2)j(j+1)}(2j+1)P_j({\bm x}{\bm\cdot}{\bm z})
\end{equation}
where $f_{\bm z}({\bm x})=\langle{\bm x}|{\bm z}\rangle$, and $|{\bm x}\rangle$ are
the common eigenvectors of the position operators $X_i$ spanning the coordinate
representation.  Namely, the probability density $p_{\bm z}({\bm x})=
|f_{\bm z}({\bm x})|^2/\|f_{\bm z}\|^2$, where the squared norm $\|f_{\bm z}\|^2
\equiv\langle{\bm z}|{\bm z}\rangle$ of the coherent state given by (15) is
\begin{equation}
\|f_{\bm z}\|^2 =
\sum_{j=0}^{\infty}e^{-j(j+1)}(2j+1)P_j(|{\bm z}|^2),
\end{equation}
where $|{\bm z}|^2={\bm z}^*{\bm\cdot}{\bm z}=\sum_{i=1}^3|z_i|^2$, is peaked at 
${\bm x}={\bm y}$ and ${\bm z}=\cosh|{\bm l}|\,{\bm y}+{\rm i}
(\sinh|{\bm l}|/|{\bm l}|){\bm l}\times {\bm y}$.
\section{Quantum beats}
Consider now a free particle on a sphere.  For the sake of simplicity we assume
that the particle has a unit mass and it moves in a unit sphere.  Clearly the
quantum Hamiltonian is given by
\begin{equation}
H=\frac{1}{2}{\bm J}^2.
\end{equation}
As with standard coherent states of harmonic oscillator the discussed coherent
states for the quantum mechanics on a sphere are not stable with respect to the
free evolution.  This can be demonstrated easily for the coherent state
$|{\bm e}_3\rangle$ given by (13) using first equation of (3a) and the relation
\cite{2}
\begin{equation}
\begin{split}
Z_3|j,0\rangle=e^{-j-1}\frac{j+1}{\sqrt{(2j+1)(2j+3)}}|j+1,0\rangle\\
{}+e^j\frac{j}{\sqrt{(2j-1)(2j+1)}}|j-1,0\rangle.
\end{split}
\end{equation}
Therefore, in view of (12) none of the coherent states is stable.  Nevertheless,
in opposition to coherent states for a particle on a plane the corresponding wave
packet (20) do not spread.  Indeed, we have
\begin{equation}
{\bm J}^2P_j({\bm x}{\bm\cdot}{\bm z})=j(j+1)P_j({\bm
x}{\bm\cdot}{\bm z}),
\end{equation}
where ${\bm J}^2=-({\bm x}\times{\bm\nabla})^2$.  Hence taking into
account (20) we find that the time-dependent coherent state in the coordinate 
representation is
\begin{equation}
\begin{split}
&f_{\bm z}({\bm x},t)=e^{-{\rm i}t{\bm J}^2/2}f_{\bm z}({\bm
x})\\
&=\frac{1}{\sqrt{4\pi}}\sum_{j=0}^{\infty}e^{-(1/2)j(j+1)(1+{\rm i}t)}
(2j+1)P_j({\bm x}{\bm\cdot}{\bm z}).
\end{split}
\end{equation}
It thus appears that $f_{\bm z}({\bm x},t)$ is $2\pi$-periodic
function of time.  From (25) and (21) it follows immediately that the
probability density for the coordinates at time $t$ is
\begin{equation}
p_{\bm z}({\bm x},t)=\frac{1}{4\pi}\frac{\big|\sum_{j=0}^{\infty}
e^{-(1/2)j(j+1)(1+{\rm i}t)}(2j+1)P_j({\bm x}{\bm\cdot}{\bm z})\big|^2}
{\sum_{j=0}^{\infty}e^{-j(j+1)}(2j+1)P_j(|{\bm z}|^2)}.
\end{equation}
The probability density (26) is a periodic function of time with period $2\pi$.
Thus it turns out that the wave packets on a sphere referring to coherent states
do not spread but rather resemble maintaining their shape solitons.  As a result
of oscillations of the probability density an interesting phenomenon takes place
that can be regarded as quantum beats on a sphere.  More precisely, consider the
following natural counterparts of the classical spherical coordinates for a free
particle on a sphere
\begin{equation}
\vartheta(t) = {\rm arccos}(e^{1/4}\langle X_3(t)\rangle_{\bm z}),
\end{equation}
where $\langle A\rangle_{\bm z}=\langle{\bm z}|A|{\bm z}\rangle/\langle
{\bm z}|{\bm z}\rangle$ and the use was made of the relation (19), and
\begin{equation}
\phi(t) = {\rm Arg}\langle X_+(t)\rangle_{\bm z},\qquad \mod 2\pi,
\end{equation}
where ${\bm X}(t)=e^{{\rm i}t{\bm J}^2/2}{\bm X}e^{-{\rm i}t{\bm J}^2/2}$,
and ${\bm z}=\cosh|{\bm l}|\,\overline{{\bm x}}+{\rm i}(\sinh|{\bm l}|/
|{\bm l}|){\bm l}\times \overline{{\bm x}}$, so $\overline{{\bm x}}$ corresponds
to the position and ${\bm l}$ to the angular momentum of a particle.  We point
out that correctness of the formula corresponding to (28) was demonstrated in the
case of the quantum mechanics on a circle \cite{4}.  The explicit formulas for the
expectation values of $X_3(t)$ and $X_+(t)$ that can be derived with the help of
(7), (3a), (5) and (14) are too complicated to reproduce them herein.  From numerical
calculations it follows that whenever the condition $l_3\approx|{\bm l}|$ holds then
the dynamics of $\vartheta(t)$ showing amplitude modulation is similar to well known
acoustical beats (see Fig.\ 1).  Clearly, the condition $l_3\approx|{\bm l}|$ is a
counterpart of the requirement of slightly different frequencies of two waves whose
superposition is the ``beat" wave.  The values of $\vartheta(t)$ oscillate around
$\overline{\theta}$ marking the coherent states via $\overline{{\bm x}}=(\sin\overline
{\theta}\cos\overline{\varphi},\sin\overline{\theta}\sin\overline{\varphi},
\cos\overline{\theta})$ (see (16)).  The minimum in the amplitude of oscillations of
$\vartheta(t)$ is reached at $t=t_*=(2k+1)\pi$, where $k$ is integer, so the period 
of beats is $2\pi$.  The dynamics of $\phi(t)$ resembles the classical one in the 
limit $l_3\to|{\bm l}|$, that is the uniform motion in a circle.  Indeed, the plot
of $\phi(t)$ is piecewise linear with constant slope for $l_3=j$, $|{\bm l}|=
\sqrt{j(j+1)}$, and $j\ge10$.  The only exception is $t=t_*$ when the amplitude
of $\vartheta(t)$ is minimal.  Namely at $t=t_*$ we have two kinds of behaviour
of the angle $\phi(t)$ shown in Fig.\ 2.  First is the simple pulse and the second
one the oscillation around $\phi=\overline{\varphi}$.  Interestingly, these two
types of graphs of the function $\phi(t)$ at $t=t_*$ occur alternately i.e.\ for
$j=10$ we have oscillation, for $j=11$ we have a pulse, for $j=12$ an oscillation
and so on.  It is worth mentioning that that the distinguished role of $t=t_*$ is
also present in the behavior of the probability density (26) having at $t=t_*$
the saddle point (see Fig.\ 3).  As a result of beats the trajectory on a sphere
parametrized as $(\sin\vartheta(t)\cos\phi(t),\sin\vartheta(t)\sin\phi(t),
\cos\vartheta(t))$ is not a grand circle as in the classical case but resembles 
the family of grand circles possessing a common diameter (see Fig.\ 4).
\begin{figure*}
\centering
\includegraphics[scale=1]{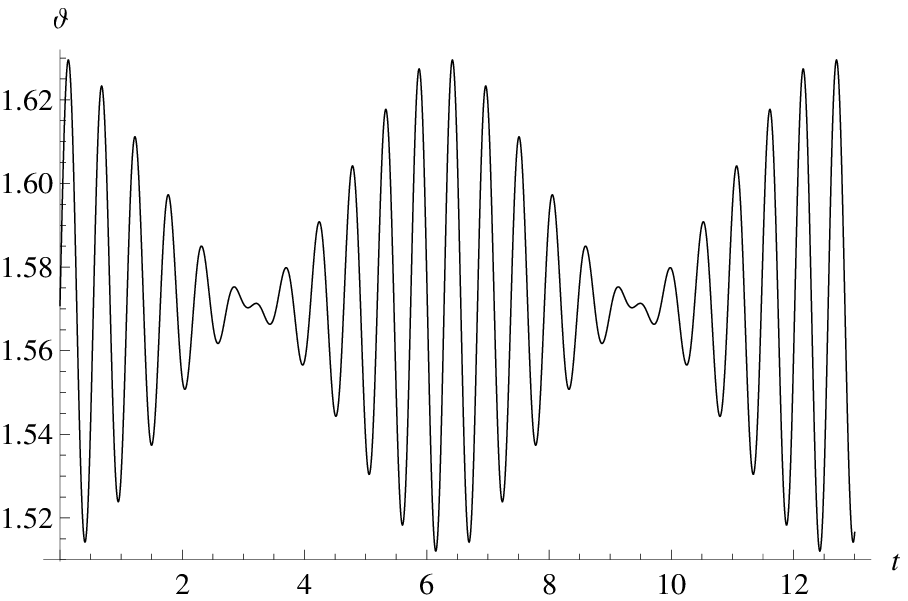}
\caption{The evolution of the counterpart $\vartheta$ of the classical angle
specifying the parallel on a sphere given by (27).  The parameters of the coherent
state $|{\bm z}\rangle$ with ${\bm z}=\cosh|{\bm l}|\,\overline{{\bm x}}+{\rm i}
(\sinh|{\bm l}|/|{\bm l}|){\bm l}\times \overline{{\bm x}}$, where 
$\overline{{\bm x}}=(\sin\overline{\theta}\cos\overline{\varphi},\sin\overline
{\theta}\sin\overline{\varphi},\cos\overline{\theta})$ are $\overline{\theta}=\pi/2$
and $\overline{\varphi}=0$.  Such values of the parameters imply $\alpha=
{\rm arccos}\,l_3/|{\bm l}|$ (see (17)).  We set $l_3=j$, and $|{\bm l}|=
\sqrt{j(j+1)}$, where $j=11$, so $l_3/|{\bm l}|=0.957$.}
\end{figure*}
\begin{figure*}
\centering
\begin{tabular}{c}
\includegraphics[scale=1]{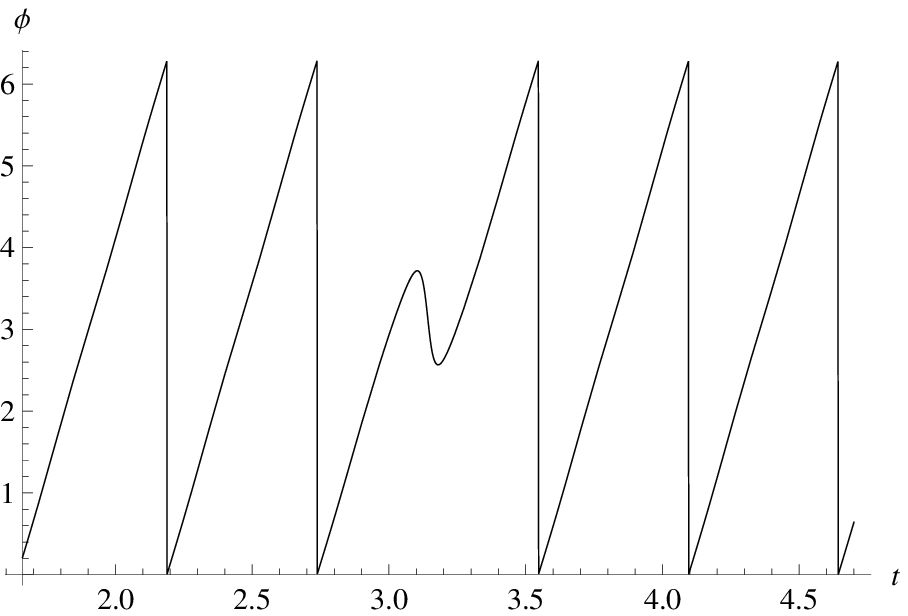}\\
\includegraphics[scale=1]{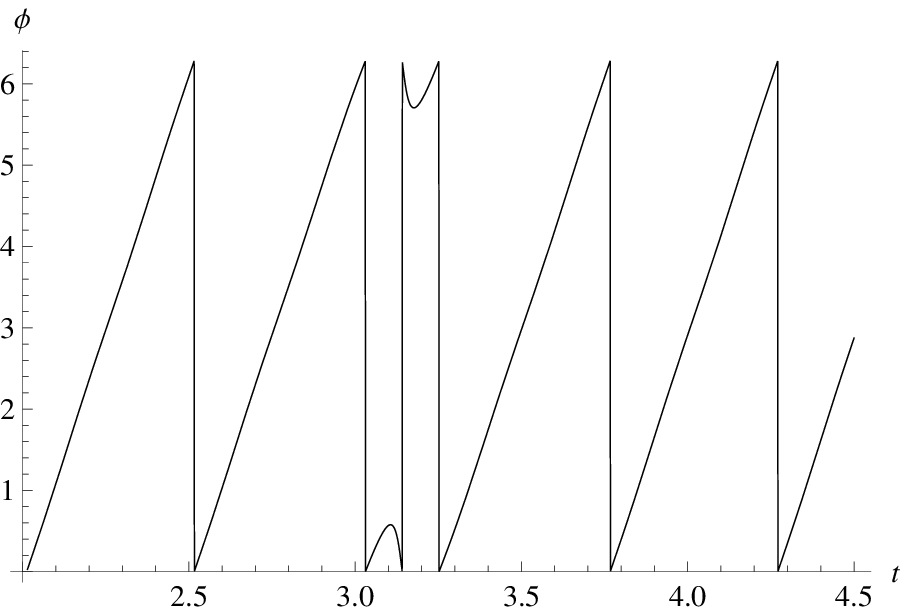}
\end{tabular}
\caption{The time-development of the counterpart $\phi$ of the classical angle
fixing the meridian on a sphere defined by (28).  Top: the parametrs the same as
in Fig.\ 1.  Bottom: the parameters the same as in Fig.\ 1 except of $j=12$.}
\end{figure*}
\begin{figure*}
\centering
\includegraphics[scale=1]{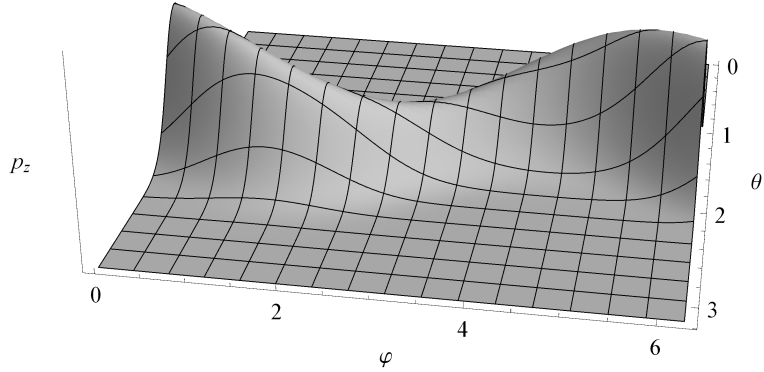}
\caption{The probability density in the case of the free evolution given by (26).
The parameters are the same as in Fig.\ 1.}
\end{figure*}
\begin{figure*}
\centering
\includegraphics[scale=1]{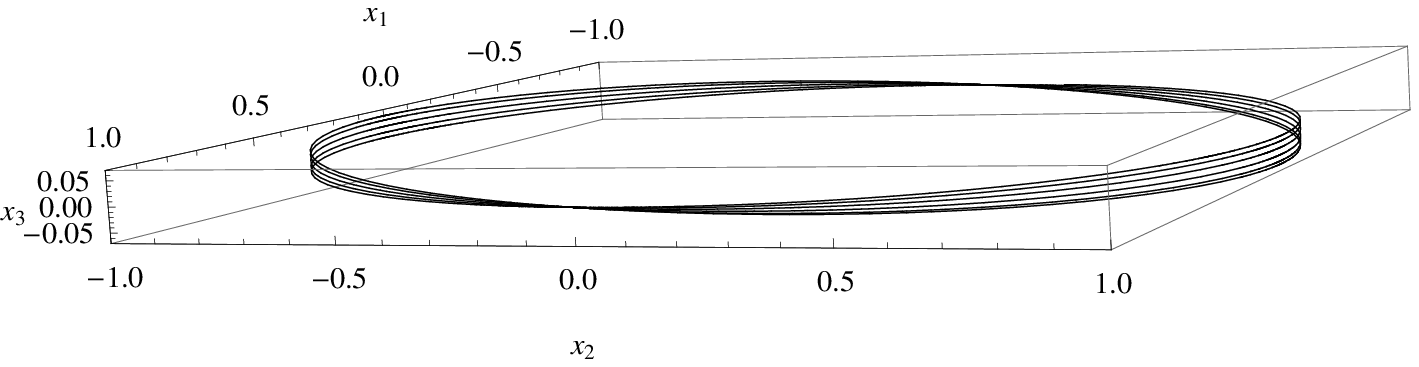}
\caption{The trajectory on a sphere parametrized as $(\sin\vartheta(t)\cos\phi(t),
\sin\vartheta(t)\sin\phi(t),\cos\vartheta(t))$, where $\vartheta(t)$ and $\phi(t)$
are the same as in Fig.\ 1 and Fig.\ 2 (top), respectively.}
\end{figure*}

It should be noted that the discussed exotic dynamics of the rigid rotor is
related to the free dynamics on a sphere that does not preserve the coherence.
In fact, the coherent states for the quantum mechanics on a sphere are stable
with respect to the evolution generated by the Hamiltonian
\begin{equation}
H_0={\bm\omega}{\bm\cdot}{\bm J},
\end{equation}
where ${\bm\omega}$ is a constant vector.  The stability of the coherent states
is an immediate consequence of the identity \cite{2}
\begin{equation}
\begin{split}
&e^{{\rm i}t{\bm\omega}{\bm\cdot}{\bm J}}{\bm Z}e^{-{\rm i}t{\bm\omega}{\bm\cdot}
{\bm J}}\\
&=\cos|{\bm\omega}|t\,{\bm Z}+\frac{\sin|{\bm\omega}|t}
{|{\bm\omega}|}
{\bm\omega}\times{\bm Z}+\frac{1-\cos|{\bm\omega}|t}{{\bm\omega}^2}{\bm\omega}
({\bm\omega}{\bm\cdot}{\bm Z})
\end{split}
\end{equation}
implying via (8)
\begin{equation}
{\bm Z}|{\bm z},t\rangle = {\bm z}(t)|{\bm z},t\rangle,
\end{equation}
where $|{\bm z},t\rangle=e^{-{\rm i}t{\bm\omega}{\bm\cdot}{\bm J}}|{\bm z}\rangle$
and ${\bm z}(t)$ is given by
\begin{equation}
{\bm z}(t)=\cos|{\bm\omega}|t\,{\bm z}+\frac{\sin|{\bm\omega}|t}
{|{\bm\omega}|}
{\bm\omega}\times{\bm z}+\frac{1-\cos|{\bm\omega}|t}{{\bm\omega}^2}{\bm\omega}
({\bm\omega}{\bm\cdot}{\bm z}).
\end{equation}
In order to compare the quantum dynamics in the case of the free evolution discussed
previously and that given by (28) we now restrict to ${\bm\omega}=(0,0,\omega_3)$, so
$H_0=\omega_3J_3$.  Evidently, with such ${\bm\omega}$ there exists a classical
solution corresponding to the uniform motion in the equator.  Using the identities
\begin{equation}
e^{{\rm i}t\omega_3J_3}X_3e^{-{\rm i}t\omega_3J_3}=X_3,\quad
e^{{\rm i}t\omega_3J_3}X_+e^{-{\rm i}t\omega_3J_3}=
e^{{\rm i}\omega_3t}X_+
\end{equation}
we find that relations (27) and (28) take the form
\begin{eqnarray}
\vartheta(t) &=& {\rm arccos}(e^{1/4}\langle X_3\rangle_{\bm z})={\rm const},\\
\phi(t) &=& {\rm Arg}\langle X_+\rangle_{\bm z}+\omega_3t,\qquad \mod 2\pi.
\end{eqnarray}
It thus appears that $\vartheta(t)$ and $\phi(t)$ follow the classical uniform circular
motion on a sphere.  In view of (19) for $\overline{\theta}=\pi/2$ and $\overline{\varphi}=0$
utilized in the case of the free evolution, where $\overline{{\bm x}}=(\sin\overline
{\theta}\cos\overline{\varphi},\sin\overline{\theta}\sin\overline{\varphi},
\cos\overline{\theta})$ is a classical position marking the coherent state, the motion
takes place in the equator.  Bearing in mind the behavior at $t=t_*$ analyzed in the
case of the free evolution we now discuss the probability density for the coordinates
corresponding to the Hamiltonian $H_0=\omega_3J_3$.  From (20) and (31) it follows
that the probability density is
\begin{equation}
p_{\bm z}({\bm x},t)=\frac{1}{4\pi}\frac{\big|\sum_{j=0}^{\infty}
e^{-(1/2)j(j+1)}(2j+1)P_j({\bm x}{\bm\cdot}{\bm z}(t))\big|^2}
{\sum_{j=0}^{\infty}e^{-j(j+1)}(2j+1)P_j(|{\bm z}|^2)},
\end{equation}
where ${\bm z}(t)$ is given by (32) with ${\bm\omega}=(0,0,\omega_3)$.  The probability 
density (36) is periodic function of time with period $2\pi/\omega_3$.  In opposition
to (26) it shows maximum and has no saddle points.

In summary, the rigid rotor in the coherent state shows unusual behavior that can be
interpreted as quantum beats.  It seems plausible to relate these beats to quantum
interference.  However, the concrete scenario is not clear.  An intriguing point is the
correlation of behavior of the rigid rotor with saddle points of the probability density.
The authors do not know any example of analogous relationship in the literature.
\section*{Acknowledgements}
This work was supported by the grant N202 205738 from the National Science Centre.

\end{document}